\documentclass[preprint,pra,superscriptaddress,showpacs,amsmath]{revtex4}

\usepackage{graphicx}

\begin{document}

\title
{Quantum optical measurements in ultracold gases:

macroscopic Bose-Einstein condensates}

\author{I. B. Mekhov}
\email{Mekhov@yahoo.com}
\affiliation{Institut f\"ur Theoretische Physik, Universit\"at
Innsbruck, Innsbruck, Austria} \affiliation{St. Petersburg State
University, Faculty of Physics, St. Petersburg, Russia}

\author{H. Ritsch}
\affiliation{Institut f\"ur Theoretische Physik, Universit\"at
Innsbruck, Innsbruck, Austria}

\begin{abstract}
We consider an ultracold quantum degenerate gas in an optical
lattice inside a cavity. This system represents a simple but key
model for "quantum optics with quantum gases," where a quantum
description of both light and atomic motion is equally important.
Due to the dynamical entanglement of atomic motion and light, the
measurement of light affects the many-body atomic state as well. The
conditional atomic dynamics can be described using the Quantum Monte
Carlo Wave Function Simulation method. In this paper, we emphasize
how this usually complicated numerical procedure can be reduced to
an analytical solution after some assumptions and approximations
valid for macroscopic Bose-Einstein condensates (BEC) with large
atom numbers. The theory can be applied for lattices with both low
filling factors (e.g. one atom per lattice site in average) and very
high filling factors (e.g. a BEC in a double-well potential). The
purity of the resulting multipartite entangled atomic state is
analyzed.
\end{abstract}

\pacs{03.75.Lm, 42.50.-p, 05.30.Jp, 32.80.Pj}

\maketitle

\newpage

\section{Introduction}

Since the first observation of the Bose-Einstein condensation (BEC)
in 1995, physics of ultracold quantum gases has become a well
established field considering various collective quantum states of
bosonic and fermionic atoms trapped in optical potentials. However,
in the majority of both theoretical and experimental works, the role
of light is reduced to a classical axillary tool for creating
fascinating quantum atomic states. On the other hand, recent
experimental achievements, where a quantum gas was loaded into a
typical quantum optical setup (high-Q cavity)
\cite{Exp1,Exp2,Exp3,Exp4,Exp5}, provides a challenge to develop a
theory of novel phenomena, where the quantizations of both light and
atomic motion play equally important roles. Recently, we have
contributed to this filed
\cite{PRL05,NatPh,PRL07,PRA07,EPJD,PRL09,PRA09,LasPhys09,Andras09,HashemArxiv},
which has stimulated further theoretical research in other groups as
well
\cite{otherTh1,otherTh2,otherTh3,otherTh4,otherTh5,otherTh6,otherTh7,otherTh8,otherTh9,otherTh10,otherTh11,otherTh12,otherTh13,otherTh14,otherTh15,otherTh16}.

At present, the main procedure to measure the properties of
ultracold atoms is the time-of-flight method, where the trapping
potential is switched off and the interference of matter waves is
observed. This method is completely destructive as the preparation
of a new atomic system is necessary for each measurement. As we have
shown in \cite{NatPh,PRL07,PRA07}, the light scattering provides a
much less destructive method to measure atomic properties, i. e.,
the quantum nondemolition (QND) measurement scheme. Light scattering
does not destroy the atomic system, hence, many consecutive
measurements of light can be done using the same atomic sample
without preparing a new one.

However, as Quantum Mechanics states, any measurement affects the
quantum state. So, even if the QND measurement does not destroy the
atomic system, it still changes the quantum states of atoms. In
Refs. \cite{NatPh,PRL07,PRA07, LasPhys09} we presented some
relations between the expectation values of the atomic and light
variables. Essentially, one needs many measurements to obtain the
expectation value of some quantum variable. Thus, even for optical
measurements, one needs to prepare the initial state several times
(in contrast to the time-of-flight schemes, the initial state can be
prepared with the same atoms).

In Refs. \cite{PRL09,PRA09}, we put a different question: the
continuous measurement of light scattered from the atoms was
considered. Due to the light-atom entanglement, the measurement of
light affects the many-body atomic state as well. Thus, the quantum
back-action of light measurement on the atomic state was analyzed.
The conditional dynamics of the atomic state due to the light
measurement was presented. In contrast to calculating the
expectation values, such an approach gives the evolution of a system
at a single quantum trajectory. This evolution should be first seen
by experimentalists before they do multiple measurements and average
them. Moreover, this measurement can be used as a method to prepare
particular atomic multipartite entangled states thanks to the
quantum measurement back-action \cite{PRL09,PRA09}.

A standard procedure to analyze the conditional evolution of the
atomic system observing light is the Quantum Monte Carlo Wave
Function (QMCWF) simulation method. Usually, it is a basis for
numerical simulations of the quantum dynamics in open dissipative
systems. In this paper, we show how this usually complicated
numerical procedure can be reduced to a simple analytical solution
after some assumptions and approximations valid for macroscopic
Bose-Einstein condensates (BEC) with large atom numbers. The purity
of the resulting multipartite entangled atomic state is analyzed.

\section{Theoretical model and quantum measurements}

\begin{figure} \scalebox{0.8}[0.8]{\includegraphics{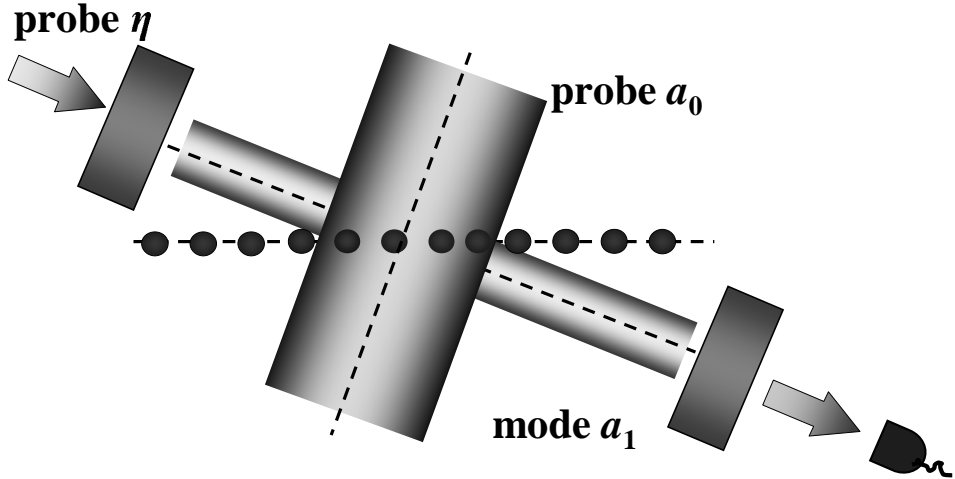}}
\caption{Setup. A lattice is illuminated by the transverse probe
$a_0$ and probe through a mirror $\eta$. The photodetector measures
photons of the cavity mode $a_1$ leaking out of the cavity. Due to
the quantum back-action, the light measurement leads to the
modification of the atomic quantum state.}
\end{figure}

We consider the model presented in Refs. \cite{PRL09,PRA09}. Using
the approximation of macroscopic atomic ensemble with the large atom
number, which is relevant for present experiments, will enable us to
obtain a simple analytical solution in the next section.

We consider (cf. Fig.~1) $N$ ultracold atoms in an optical lattice
of $M$ sites formed by strong off-resonant laser beams. A region of
$K\le M$ sites is also illuminated by a weak external probe, which
is scattered into a cavity. Alternatively, this region is
illuminated by the cavity field appearing due to the presence of the
probe through the cavity mirror.

We use the open system approach for counting photons leaking the
cavity of decay rate $\kappa$. When a photon is detected, the jump
operator (the cavity photon annihilation operator $a_1$) is applied
to the quantum state: $|\Psi_c(t)\rangle \rightarrow
a_1|\Psi_c(t)\rangle$. Between the counts, the system evolves with a
non-Hermitian Hamiltonian. Such an evolution gives a quantum
trajectory for $|\Psi_c(t)\rangle$ conditioned on the detection of
photons.

The expression for the initial motional state of atoms reads
\begin{eqnarray}\label{1}
|\Psi(0)\rangle =\sum_{q}c_q^0 |q_1,..,q_M\rangle,
\end{eqnarray}
which is a superposition of Fock states reflecting all possible
classical configurations $q=\{q_1,..,q_M\}$ of $N$ atoms at $M$
sites, where $q_j$ is the atom number at the site $j$. As we have
shown in Refs. \cite{PRL09,PRA09}, the solution for conditional wave
function takes physically transparent form, if the following
approximations are used: atomic tunneling is much slower than light
dynamics, the probe waves are in the coherent state, the first
photon is detected after the time $1/\kappa$. The conditional state
after the time $t$ and $m$ photocounts is given by the quantum
superposition of solutions corresponding to the atomic Fock states
in Eq.~(\ref{1}):
\begin{eqnarray}
|\Psi_c(m,t)\rangle =\frac{1}{F(t)}\sum_{q}\alpha_q^m e^{\Phi_q(t)}
c_q^0 |q_1,...,q_M\rangle|\alpha_q\rangle, \label{2}\\
\alpha_q=\frac{\eta-iU_{10} a_0D^q_{10}}{i(U_{11}
D^q_{11}-\Delta_p)+\kappa}, \label{3}\\
\Phi_q(t)=-|\alpha_q|^2\kappa t+(\eta\alpha^*_q-iU_{10}
a_0D^q_{10}\alpha^*_q-\text{c.c.})t/2, \label{4}
\end{eqnarray}
where $\alpha_q$ is the cavity light amplitude corresponding to the
classical configuration $q$. It is simply given by the Lorentz
function (\ref{3}) well-known from classical optics, where $a_0$ is
the external probe amplitude, $\eta$ is the amplitude of the probe
through a mirror; $U_{lm}=g_lg_m/\Delta_a$ ($l,m=0,1$), where
$g_{1,0}$ are the atom-light coupling constants, $\Delta_a=\omega_1
- \omega_a$ is the cavity-atom detuning; $\Delta_p=\omega_p -
\omega_1$ is the probe-cavity detuning. $D^q_{lm}=
\sum_{j=1}^K{u_l^*({\bf r}_j)u_m({\bf r}_j)q_j}$ are the
probe-cavity coupling coefficient and dispersive frequency shift
that sums contributions from all illuminated atoms with prefactors
given by the light mode functions $u_{0,1}({\bf r})$. Except the
prefactors associated with $m$ photodetections $\alpha_q^m$, the
components of quantum superposition in Eq.~(\ref{2}) acquires the
phases contained in $\Phi_q(t)$ (\ref{4}). $F(t)$ is the
normalization coefficient.

For several particular cases, the solution for the time-dependent
probability distribution of atoms corresponding to the state
(\ref{2}) can be simplified further \cite{PRL09,PRA09}. If the
probe, cavity, and lattice satisfy the condition of the diffraction
maximum for light scattering, the probability to find the atom
number $0<z<N$ in the lattice region of $K$ sites is given by
\begin{eqnarray}\label{5}
p(z,m,t)=z^{2m}e^{-z^2\tau}p_0(z)/\tilde{F}^2, \\
\tilde{F}^2=\sum_z z^{2m}e^{-z^2\tau}p_0(z), \nonumber
\end{eqnarray}
with $\tau=2|C|^2\kappa t$, $C=iU_{10} a_0/(i\Delta_p-\kappa)$,
$p_0(z)$ is the initial distribution, and $\tilde{F}$ provides the
normalization. The light amplitude corresponding to the atom number
$z$ is $\alpha_z=Cz$.

If the condition of a diffraction minimum is satisfied, the
probability to find the atom number difference between the odd and
even sites in the lattice region of $K=M$ sites is given by the same
Eq.~(\ref{5}), but with a different meaning of the statistical
variable $-N<z<N$.

When the time progresses, both $m$ and $t$ increase with an
essentially probabilistic relation between them. The Quantum Monte
Carlo method, which establishes such a relation thus giving a
quantum trajectory, consists in the following. The evolution is
split into small time intervals $\delta t_i$. In each time step, the
conditional photon number is calculated using the probability
distribution (\ref{5}):
\begin{eqnarray}\label{6}
\langle a^\dag_1a_1\rangle_c(t)=|C|^2 \sum_z z^2p(z,m,t),
\end{eqnarray}
which is proportional to the second moment of $p(z,m,t)$. The
probability of the next, $(m+1)$th, photocount within this time
interval $P_{m+1}=2\kappa \langle a^\dag_1a_1\rangle_c \delta t_i$
is then compared with a random number $0<\epsilon_i<1$ generated in
advance, thus, deciding whether the detection (if $2\kappa \langle
a^\dag_1a_1\rangle_c \delta t_i>\epsilon_i$) or no-count process
(otherwise) has happened.

\section{Macroscopic quantum gases}

In this section, we consider a case, where the initial atomic state
is a macroscopic superfluid (SF) with the atom number $N\gg1$. Note,
that the total number of lattice sites $M$ and the number of sites
illuminated $K$ can be any. Thus, the theory presented below can be
applied for lattices with both low filling factors (e.g. one atom
per lattice site in average) and very high filling factors (e.g. a
BEC in a double-well potential).

We will start with the case of a diffraction minimum. As shown in
Refs. \cite{PRL09,PRA09}, the probability function (\ref{5}) to find
the atom number difference $z$ between the odd and even sites
shrinks to a doublet with the peaks at $z_{1,2}=\sqrt{m/\tau}$,
which corresponds to the generation of the Schr{\"o}dinger cat
state.

For the SF state the probability to find the atom number at odd (or
even) sites $\tilde{z}$ [$\tilde{z}=(z+N)/2$ as the atom number
difference is $z$ and the total atom number is $N$] is given by the
binomial distribution
\begin{eqnarray}\label{7}
p_\text{SF}(\tilde{z})=\frac{N!}{\tilde{z}!(N-\tilde{z})!}
\left(\frac{Q}{M}\right)^{\tilde{z}}
 \left(1-\frac{Q}{M}\right)^{N-\tilde{z}},
\end{eqnarray}
where $Q$ is the number of odd (or even) sites. For even $M$,
$Q=M/2$ and Eq.~(\ref{7}) simplifies. For a lattice with the large
atom number $N\gg1$, this binomial distribution can be approximated
by a Gaussian function. Changing the variable as $z=2\tilde{z}-N$ we
obtain the Gaussian function for the probability to find the atom
number difference $z$:
\begin{eqnarray}\label{8}
p_\text{SF}(z)=\frac{1}{\sqrt{2\pi}\sigma}e^{-\frac{z^2}{2\sigma^2}}
\end{eqnarray}
with the zero mean $z$ and $\sigma=\sqrt{N}$ giving the full width
at half maximum (FWHM) $2\sigma\sqrt{2\ln 2}$. The variance of the
atom number difference in the SF state is $\sigma^2=N$.

For the large atom number, the summations in Eqs.~(\ref{5}) and
(\ref{6}) can be replaced by the integrals over all $-\infty
<z<\infty$, which gives the following probability of the next
photocount:
\begin{eqnarray}\label{9}
P_{m+1}= \frac{\int_{-\infty}^\infty{z^{2m+2}e^{-z^2\tau}
e^{-\frac{z^2}{2\sigma^2}}dz}}{\int_{-\infty}^\infty{z^{2m}e^{-z^2\tau}
e^{-\frac{z^2}{2\sigma^2}}dz}}\delta\tau_i,
\end{eqnarray}
where $\delta\tau_i=2|C|^2\kappa\delta t_i$. Taking into account the
following relation \cite{Gradstein}:
\begin{eqnarray}\label{10}
\int_{0}^\infty{x^{2n}e^{-px^2}dx}=\frac{(2n-1)!!}{2(2p)^n}\sqrt{\frac{\pi}{p}},
\end{eqnarray}
the integrals can be calculated and the probability of the next
photocount reads
\begin{eqnarray}\label{11}
P_{m+1}=\frac{m+1/2}{\tau+1/\sigma^2}\delta\tau_i.
\end{eqnarray}

Thus, we see that the Quantum Monte Carlo method, which is usually
expected to be a basis for hard numerical simulations, has reduced
to an extremely simple form. After splitting our time axis into
intervals $\delta\tau_i$ and generating the random numbers
$0<\epsilon_i<1$, one has simply to substitute the current time
$\tau$ and the photocount number $m$ in the trivial algebraic
expression (\ref{11}) and realize if the next photocount happened or
not. Proceeding this way one establishes the relation between the
photocount number and time $m(\tau)$ at the quantum trajectory
corresponding to the generated set of random numbers $\epsilon_i$.
Knowing the relation between $m$ and $\tau$, one can calculate
various conditional expectation values using Eq.~(\ref{10}) and the
complementary expression
\begin{eqnarray}\label{12}
\int_{0}^\infty{x^{2n+1}e^{-px^2}dx}=\frac{n!}{2p^{n+1}}.
\end{eqnarray}

We now switch to the case of the light detection at the direction of
a diffraction maximum. As shown in Refs. \cite{PRL09,PRA09}, in this
case one measures the atom number $z$ in the lattice region of $K$
illuminated sites.

For the initial SF state the probability to find the atom number $z$
at the lattice region of $K$ sites is given by the binomial
distribution
\begin{eqnarray}\label{13}
p_\text{SF}(z)=\frac{N!}{z!(N-z)!}\left(\frac{K}{M}\right)^z
\left(1-\frac{K}{M}\right)^{N-z}.
\end{eqnarray}
For a lattice with the large atom number $N\gg1$, it can be
approximated as a Gaussian distribution
\begin{eqnarray}\label{14}
p_\text{SF}(z)=\frac{1}{\sqrt{2\pi}\sigma}e^{-\frac{(z-z_0)^2}{2\sigma^2}}
\end{eqnarray}
with the mean atom number $z_0=NK/M$ and
$\sigma=\sqrt{N(K/M)(1-K/M)}$ giving FWHM $2\sigma\sqrt{2\ln 2}$.
The atom number variance in the SF state is $\sigma^2$.

Similarly to the case of a diffraction minimum, for the large atom
number, the summations in Eqs.~(\ref{5}) and (\ref{6}) can be
replaced by the integrals over $-\infty<z<\infty$, which gives the
following probability of the next photocount:
\begin{eqnarray}\label{15}
P_{m+1}= \frac{\int_{0}^\infty{z^{2m+2}e^{-z^2\tau}
e^{-\frac{(z-z_0)^2}{2\sigma^2}}dz}}{\int_{0}^\infty{z^{2m}e^{-z^2\tau}
e^{-\frac{(z-z_0)^2}{2\sigma^2}}dz}}\delta\tau_i.
\end{eqnarray}
Taking into account the following relation \cite{Gradstein}:
\begin{eqnarray}\label{16}
\int_{-\infty}^\infty{x^{n}e^{-px^2+2qx}dx}=\frac{1}{2^{n-1}p}\sqrt{\frac{\pi}{p}}
\frac{d^{n-1}}{dq^{n-1}}\left(qe^\frac{q^2}{p}\right) \\\nonumber
=n!e^\frac{q^2}{p}\sqrt{\frac{\pi}{p}}\left(\frac{q}{p}\right)^n\sum_{k=0}^{E(n/2)}
\frac{1}{(n-2k)!k!}\left(\frac{p}{4q^2}\right)^k,
\end{eqnarray}
where $E(n/2)$ is the integer part of $n/2$, the integrals can be
calculated and the probability of the next photocount reads
\begin{eqnarray}\label{17}
P_{m+1}=(2m+1)(2m+2)a^2\sum_{k=0}^{m+1}
\frac{b^k}{(2m+2-2k)!k!}/\sum_{k=0}^{m} \frac{b^k}{(2m-2k)!k!},
\end{eqnarray}
where the parameters are
\begin{eqnarray}
a=\frac{z_0}{2\sigma^2(\tau+1/(2\sigma^2))}=\frac{1}{2(1-K/M)(\tau+1/(2\sigma^2))},\nonumber\\
b=\frac{\tau+1/(2\sigma^2)}{z_0^2}\sigma^4=(\tau+1/(2\sigma^2))\left(1-\frac{K}{M}\right)^2.\nonumber
\end{eqnarray}
Expression (\ref{17}) is more complicated than Eq.~(\ref{11}) for
the diffraction minimum. However, it is also very simple as it
includes only summation over the photocount number. Thus, we were
able to replace the summation over the atom number, which can rich
the values of $N=10^5$, and even the numerical integration over the
atom number. For the far off-resonant interaction considered here,
the number of photocounts $m$ will be many orders of magnitude less
than the atom number. So, the sum in Eq.~(\ref{17}) will contain
only a small number of terms.

\section{Purity of the final state}

As shown in Ref. \cite{PRA09}, under some conditions, the entangled
light-matter state can collapse to a macroscopic superposition
state, which has a form
\begin{eqnarray}\label{18}
|\Psi_c\rangle=\frac{1}{\sqrt{2}}[e^{i\gamma}|z_1\rangle|\alpha_{z_1}\rangle
+ e^{- i\gamma}|z_2\rangle|\alpha_{z_2}\rangle],
\end{eqnarray}
where $\gamma$ is some phase, $|z_{1,2}\rangle$ are the atomic Fock
states with precisely known atom numbers $z_1$ and $z_2$.
$|\alpha_{z_{1,2}}\rangle$ are the corresponding coherent states of
light such that the light amplitudes have equal absolute values, but
opposite phases: $\alpha_{z_1}=|\alpha_{z_1}|\exp{(i\varphi)}$ and
$\alpha_{z_2}=|\alpha_{z_1}|\exp{(-i\varphi)}$.

Expression (\ref{18}) gives a macroscopic superposition state, where
light and atoms are entangled. One possibility to disentangle them
is to switch off the probe and count the photons leaking out of the
cavity. In this case both $|\alpha_{z_{1}}\rangle$ and
$|\alpha_{z_{2}}\rangle$ will approach the vacuum state $|0\rangle$
and the light and matter will factorize. The difficulty of such a
method consists in the necessity to count almost all photons to get
finally an atomic state of high purity \cite{PRA09}. In this paper,
we analyze the state obtained by tracing out the light field
directly in Eq. (\ref{18}), without waiting for photon leakage. This
corresponds to a scheme, where the measurement of light is not
performed.

The trace over the light variables can be calculated using the
photon Fock basis, in which the coherent light states read
\begin{eqnarray}\label{19}
|\alpha_{z_{1,2}}\rangle=e^{-|\alpha_{z_1}|^2/2}\sum_{n=0}^\infty\frac{|\alpha_{z_{1}}|^ne^{\pm
n\varphi}}{\sqrt{n!}}|n>.
\end{eqnarray}
The density matrix of the state after tracing out the light is
$\rho=\sum_{n=0}^\infty\langle n|\Psi_c\rangle \langle
\Psi_c|n\rangle$. Using the expression
\begin{eqnarray}\label{20}
\sum_{n=0}^\infty\frac{|\alpha_{z_1}|^{2n}}{n!}e^{-|\alpha_{z_1}|^2}e^{-2in\varphi}=
e^{|\alpha_{z_1}|^2(e^{-2i\varphi}-1)},
\end{eqnarray}
the density matrix takes the form
\begin{eqnarray}\label{21}
\rho=\frac{1}{2}\left(|z_1\rangle\langle z_1| + |z_2\rangle\langle
z_2|+
e^{|\alpha_{z_1}|^2(e^{2i\varphi}-1)}e^{2i\gamma}|z_1\rangle\langle
z_2|+e^{|\alpha_{z_1}|^2(e^{-2i\varphi}-1)}e^{-2i\gamma}|z_2\rangle\langle
z_1|\right)
\end{eqnarray}

The quantity characterizing how close is a mixture state to a pure
state is the so-called purity: $P=\text{Tr}(\rho^2)$. For a pure
state it is maximal and equal to 1, while for a maximally mixed
state it is minimal and equal to $1/2$ (in our case of the
two-component states). The purity of the state (\ref{21}) is given
by
\begin{eqnarray}\label{22}
P=\frac{1}{2}\left(1+e^{-4|\alpha_{z_1}|^2\sin^2\varphi}\right).
\end{eqnarray}

The purity depends on the amplitude of the coherent light states in
Eq. (\ref{18}) and the phase difference between them. In a trivial
case, where two coherent states are indistinguishable ($\varphi=0$),
the purity is maximal, $p=1$, and the state is pure (however, in
this case, $z_1=z_2$ and the state is not a macroscopically
entangled one). In non-trivial cases, where the coherent states
differ by the phase $2\varphi$, the purity decreases with increase
of the light amplitude and phase difference between them. One can
estimate the minimal possible purity as follows. In a coherent
state, the uncertainty of the $X$-quadrature is $1/2$. Thus, two
coherent states can be well distinguished, if
$|\alpha_{z_1}|\sin\varphi>1/4$. Substituting the minimal value 1/4
in Eq. (\ref{22}), one sees that the maximal purity can reach 0.89,
which is a rather high value.

\section{Conclusions}

We have considered the quantum measurement of light scattered from
an ultracold quantum gas. Due to the light-matter entanglement, the
back-action of light measurement modifies the atomic state as well.
We applied our theory developed in Refs. \cite{PRL09,PRA09} for the
case of a macroscopic BEC with a large atom number. We have shown
that the Quantum Monte Carlo simulation method gets very simple in
that case and the calculations can be carried out even analytically.
The purity of the macroscopic superposition atomic state was
analyzed and shown to be able to reach the high values.


\begin{thebibliography}{19}
\bibitem{Exp1} F.\ Brennecke, T. Donner, S. Ritter, T. Bourdel, M. Kohl, and T.
Esslinger, Nature {\bf 450}, 268 (2007).
\bibitem{Exp2} Y.\ Colombe, T. Steinmetz, G. Dubois, F. Linke, D. Hunger, and J.
Reichel, Nature {\bf 450}, 272 (2007).
\bibitem{Exp3} S.\ Slama, S. Bux, G. Krenz, C. Zimmermann, and Ph. W. Courteille,
Phys.\ Rev.\ Lett. {\bf 98}, 053603 (2007).
\bibitem{Exp4} S. Ritter, F. Brennecke, C. Guerlin, K. Baumann, T. Donner, T.
Esslinger, Appl. Phys. B {\bf 95}, 213 (2009).
\bibitem{Exp5} F. Brennecke, S. Ritter, T. Donner, T. Esslinger, Science {\bf 322},
235 (2008).
\bibitem{PRL05} C.\ Maschler and H.\ Ritsch, Phys.\ Rev.\ Lett. {\bf 95}, 260401 (2005).
\bibitem{NatPh} I.\ B.\ Mekhov, C.\ Maschler, and H.\ Ritsch, Nature Phys. {\bf 3}, 319 (2007).
\bibitem{PRL07} I.\ B.\ Mekhov, C.\ Maschler, and H.\ Ritsch, Phys.\ Rev.\ Lett. {\bf 98}, 100402 (2007).
\bibitem{PRA07} I.\ B.\ Mekhov, C.\ Maschler, and H.\ Ritsch, Phys.\ Rev.\ A {\bf 76}, 053618 (2007).
\bibitem{EPJD} C.\ Maschler, I.\ B.\ Mekhov, and H.\ Ritsch, Eur.\ Phys.\ J.\ D {\bf 46}, 545 (2008).
\bibitem{PRL09} I. B. Mekhov and H. Ritsch, Phys. Rev. Lett. {\bf 102}, 020403 (2009).
\bibitem{PRA09} I.\ B.\ Mekhov, and H.\ Ritsch, Phys.\ Rev.\ A {\bf 80}, 013604 (2009).
\bibitem{LasPhys09} I. B. Mekhov and H. Ritsch, Laser Phys. {\bf 19}, 610 (2009).
\bibitem{Andras09} A. Vukics, W. Niedenzu, and H. Ritsch, Phys. Rev. A {\bf 79}, 013828 (2009).
\bibitem{HashemArxiv} H. Zoubi and H. Ritsch, arXiv:0902.2638.
\bibitem{otherTh1} W. Chen, D. Meiser, and P. Meystre, Phys. Rev. A
{\bf 75}, 023812 (2007).
\bibitem{otherTh2} J. Larson, B. Damski, G. Morigi, and M. Lewenstein, Phys. Rev. Lett.
{\bf 100}, 050401 (2008).
\bibitem{otherTh3} J. Larson, S. Fernandez-Vidal, G. Morigi, and M. Lewenstein, New J.
Phys. {\bf 10}, 045002 (2008).
\bibitem{otherTh4} K. Eckert, O. Romero-Isart, M. Rodriguez, M. Lewenstein, E. Polzik,
and A. Sanpera, Nature Phys. {\bf 4}, 50 (2008).
\bibitem{otherTh5} A. B. Bhattacherjee, Opt. Commun. {\bf 281}, 3004 (2008).
\bibitem{otherTh6} J. M. Zhang, W. M. Liu, and D. L. Zhou, Phys. Rev. A {\bf 77},
033620 (2008).
\bibitem{otherTh7} J. M. Zhang, W. M. Liu, and D. L. Zhou, Phys. Rev. A {\bf 78},
043618 (2008).
\bibitem{otherTh8} J. Ye, J. Zhang, W. Liu, K. Zhang, Y. Li, Z.-Y. Ou, and W. Zhang,
arXiv:0812.4077.
\bibitem{otherTh9} A. B. Bhattacherjee, arXiv:0906.2624.
\bibitem{otherTh10} K. Lakomy, Z. Idziaszek, and M. Trippenbach, arXiv:0904.2927.
\bibitem{otherTh11} L. Guo, S. Chen, B. Frigan, L. You, and Y. Zhang, Phys. Rev. A {\bf
79}, 013630 (2009).
\bibitem{otherTh12} W. Chen and P. Meystre, Phys. Rev. A {\bf 79}, 043801 (2009);
\bibitem{otherTh13} W. Chen, K. Zhang, D. S. Goldbaum, M. Bhattacharya, and P. Meystre,
Phys. Rev. A {\bf 80}, 011801(R) (2009).
\bibitem{otherTh14} K. Zhang, W. Chen, P. Meystre, arXiv:0906.4143.
\bibitem{otherTh15} J. M. Zhang, S. Cui, H. Jing, D. L. Zhou, and W. M. Liu,
arXiv:0907.1200.
\bibitem{otherTh16} S. Rist, C. Menotti, and G. Morigi, arXiv:0904.0915.
\bibitem{Gradstein} L. S. Gradstein and I. M. Ryzhik, {\it Tables of Integrals, Series and
Products} (Academic Press, 1980).
\end{thebibliography}
\end{document}